\documentclass[superscriptaddress,preprint]{revtex4}
\usepackage{mathrsfs}
\usepackage{amssymb}
\usepackage[tbtags]{amsmath}
\usepackage{graphicx}
\usepackage{epsfig,graphicx,times}
\usepackage{color}
\usepackage{subfigure}

\begin{document}
\title{Application of the weak-measurement technique to study atom-vacuum interactions}
\author{M. Zhang}
\affiliation{Laboratory for Quantum Optics and Quantum Information,
Beijing Computational Science Research Center, Beijing 100084,
China} \affiliation{Department of Physics, Southwest Jiaotong
University, Chengdu 610031, China}
\author{S. Y. Zhu\footnote{syzhu@csrc.ac.cn}}
\affiliation{Laboratory for Quantum Optics and Quantum Information,
Beijing Computational Science Research Center, Beijing 100084,
China}
\date{\today}

\begin{abstract}
Quantum weak measurement has attracted much interest recently [J.
Dressel \emph{et al}.,~{\color{blue}Rev. Mod. Phys. 86, 307
(2014)]}, because it could amplify some weak signals and provide a
technique to observe nonclassical phenomena. Here, we apply this
technique to study the interaction between the free atoms and the
vacuum in a cavity. Due to the gradient field in the vacuum cavity,
the external orbital motions and the internal electronic states of
atoms can be weakly coupled via the atom-field electric-dipole
interaction. We show that, within the properly postselected internal
states, the weak atom-vacuum interaction could generate a large
change to the external motions of atoms due to the
postselection-induced weak values.
\end{abstract}

\maketitle
\section{Introduction}
The conception of quantum weak measurement was introduced by
Aharonov, Albert, and Vaidman (AAV) in 1988~\cite{AAV}. Their theory
is based on the von Neumann measurement with a very weak coupling
between two quantum systems~\cite{Jozsa}, for example, the weak
spin-orbit coupling of electrons in the Stern-Gerlach (SG) device. A
key feature of the weak measurement is that the observable quantity
(acting as the pointer) is measured in a certain subensemble, for
example, measuring the expectation value of the electrons' position
with the postselected spin state $|f\rangle$. This measurement leads
to an interesting result that the pointer has a shift proportional
to the value
\begin{equation}
A_w=\frac{\langle f|\hat{A}|i\rangle}{\langle f|i\rangle}\,,
\end{equation}
where $|i\rangle$ and  $\hat{A}$ are, respectively, the initial
state and the observable operator of the spin system. $A_w$ is the
so-called weak value. Compared to the strong measurement $\langle
i|\hat{A}|i\rangle$, the weak value provides an improved approach to
detect $\hat{A}$, and some interesting phenomena result.

Recently, the weak value has attracted much interest because it
could be arranged to amplify some weak signals~\cite{Amplification1,
Amplification2,Amplification3,Amplification4,Amplification5,Hall
Effect}. It is also used to study the foundational questions of
quantum
mechanics~\cite{three-box,C-Cat,Nature-wavefuction,Science-trajectory,Superluminal},
such as Hardy's paradox~\cite{HD}, the Leggett-Garg
inequality~\cite{LG}, Heisenberg's uncertainty
relation~\cite{Uncertain}, and the wave-particle
correlation~\cite{correlation}. Regarding the physical
implementations, most of the previous studies used the light both as
the pointer and the measured system~\cite{RMP}. There are several
interesting works implementing weak measurement using the
condensed-matter system, e.g., the quantum dot~\cite{Dot}, the
superconducting phase qubit~\cite{Superconducting}, and the
semiconducting Aharonov-Bohm interferometer~\cite{AB}. Recently,
Ref.~\cite{Atom} studied the weak measurement of a cold-atom system
based on the dynamics of spontaneous emission.

In this article, the weak measurement is applied to the system of
atom-cavity interaction. In such a system, the cavity
electrodynamics (cavity QED) have predicted many nonclassical
phenomena such as the famous vacuum Rabi
oscillation~\cite{HarocheRabi,HarocheCat,HarocheRMP} and the vacuum
Rabi splitting~\cite{splitting1983,splittingkimble,splittingNature}.
These effects concern the cavity-induced changes in the internal
electron's states of atoms. Remarkably, it has been shown that the
light in a cavity can significantly affect the atom's center-of-mass
(c.m.) motions, for example, Kapitza-Dirac
scattering~\cite{Kapitza-Dirac-vacuum,Kapitza-Dirac-0,Kapitza-Dirac-1,Kapitza-Dirac-2,Kapitza-Dirac-RMP}.
This effect is due to the atom stimulated emitting and absorbing
photon in the cavity (resulting in a momentum change in the atom).
It can be found that a vacuum cavity can also generate a similar
transverse effect of a neutral atom via the virtual excitation of a
photon. Here, we propose a weak value amplification (WVA) setup to
observe such an interesting nonclassical effect of vacuum. After the
atom-cavity interaction, we perform a single-qubit operation on the
two internal states of atoms and postselected on an internal state.
Then, we obtain a weak value; its real and imaginary parts
determine, respectively, the shifts of the average momentum and the
position of the atoms' external motions. Consequently, the
controllable weak value could be used to amplify the vacuum-induced
transverse shifts of atoms. It is shown that the present WVA could
offer some certain advantages for experimentally detecting the weak
transverse effects of atoms.

Our paper is organized as follows. In Sec.~II, we present the
vacuum-induced weak coupling between the internal and external
motions of free atoms. This coupling acts as a force to push the
neutral atoms moving transversely. In Sec. III, we get the desirable
weak value using the single-qubit operation and postselection and
use it to amplify the transverse shifts of atoms. In Sec. IV, we
discuss the physical meaning of WVA. Our conclusions are summarized
in Sec. V.

\section{The vacuum-induced coupling between the internal and external motions of free atoms}
Following the original work of AAV, we consider the weak measurement
experiment as showing in Fig.~1. The spatially coherent atoms, e.g.,
a released BEC~\cite{Kapitza-Dirac-RMP}, are injected into the
equipment through a pinhole located about the point of $(0,0,0)$.
This pinhole selects a part of the matter wave, and thus the
positional uncertainty of the selected atoms is on the order of the
size of pinhole. Hence, one can use the typical Gaussian wave-packet
to describe the spatially coherent atoms (after the pinhole). In
$x$-direction, the Gaussian state reads
\begin{equation}
|G\rangle=\int_{-\infty}^{\infty}dx \phi(x) |x\rangle\,.
\end{equation}
Where $\phi(x)=\langle
x|G\rangle=(2\pi\Delta^2)^{-1/4}\exp[-x^2/(4\Delta^2)]$ is the
probability-amplitude of position eigenstate $|x\rangle$ and
$\Delta$ describes the root-mean-square (rms) width of the
wave-packet. Of course, the state (2) can be also written as
$|G\rangle=\int_{-\infty}^{\infty}dp\, \phi(p) |p\rangle$ with the
momentum eigenstate $|p\rangle$ and the Gaussian function
$\phi(p)=\langle
p|G\rangle=[2\Delta^2/(\pi\hbar^2)]^{1/4}\exp(-\Delta^2p^2/\hbar^2)$.
For this Gaussian state, the expectation value of position is
$\langle x\rangle=0$ and its uncertainty reads $\Delta=\sqrt{\langle
x^2\rangle-\langle x\rangle^2}$. The average momentum along $x$
direction is $\langle p\rangle=0$ and its uncertainty reads
$\Delta_{p}=\sqrt{\langle p^2\rangle-\langle
p\rangle^2}=\hbar/(2\Delta)$. Physically, the uncertainty $\Delta$
(or $\Delta_p$) determines the main distribution range of particles'
positions (or momentums). Out of this range, the probability to find
the particles is negligible. Below, we study the vacuum field (in
the cavity 1) induced change on the initial wave packet $\phi(x)$
within a very short duration (i.e., the free diffraction of atom is
negligible).

\begin{figure}[tbp]
\includegraphics[width=14cm]{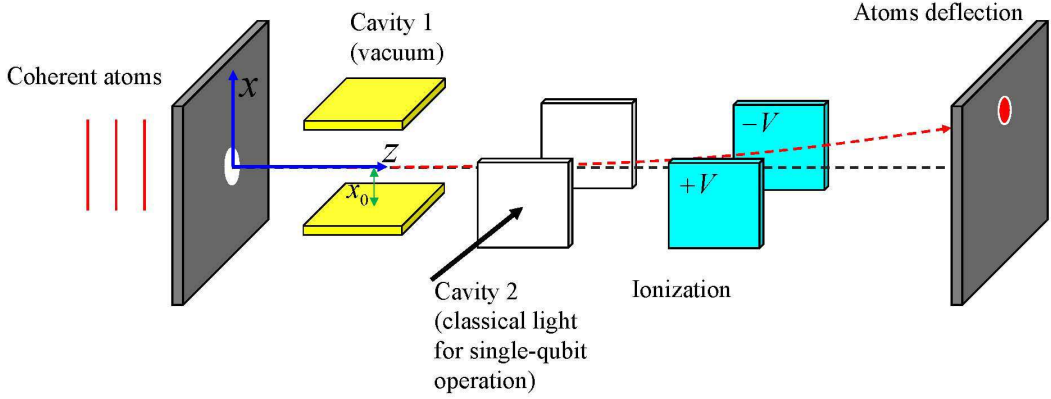}
\caption{(Color online) Sketch of the weak measurement process. The
two-levels atoms are prepared in a certain internal state
$|S_i\rangle$, and pass through a pinhole with momentum along the
$z$ direction. The vacuum field (with the $x$-directional gradient)
in cavity $1$ generates a weak coupling between the atoms' internal
states and the external $x$-directional motions. Cavity $2$, with
classical light, resonantly excites atoms and generates the
desirable single-qubit operation $\hat{U}$. The applied voltage $\pm
V$ ionizes the atoms in excited state (similar to the procedure in
the experiments of Haroche
group~\cite{HarocheRabi,HarocheCat,HarocheRMP}) and leave the ground
state atoms to be detected. In the selected ensemble of ground
state, the atoms have a shift (along $x$ direction) of the average
position on the deposition plate. This shift can be described by the
so-called weak value, which depends on the pre-selection
$|S_i\rangle$ and the single-qubit operation $\hat{U}$. }
\end{figure}

In cavity $1$, the quantized field of a mode takes the
form~\cite{Scully}
\begin{equation}
\vec{E}=\vec{\tau}E_{0}\sin (kx+kx_0)(\hat{a}^\dagger+\hat{a})
\end{equation}
which excites the incoming atoms. Here, $\vec{\tau}$, $E_{0}$ and
$k$ are respectively the polarization-vector, amplitude, and
wave-number of the standing wave (such as the first excited mode).
$\hat{a}^\dagger$ and $\hat{a}$ are respectively the creation and
annihilation operators of the corresponding cavity mode (with
frequency $\omega_c$).
We consider the microwave excitation of the two-level Rydberg atoms.
Although the orbit radius of Rydberg states are very large (about
$10^3$ atomic units~\cite{HarocheRabi,HarocheCat,HarocheRMP}), they
are far smaller than the wavelength of the microwave cavity (on the
order of centimeter). Therefore, in the atomic internal region the
driving field (3) can be regarded as uniform. Performing the dipole
approximation, the interaction between the atom and cavity field
reads
\begin{equation}
\hat{H}_{\rm int}=\hbar\Omega_{0}\sin
(kx+kx_0)(\hat{a}^\dagger+\hat{a})\hat{\sigma}_x
\end{equation}
with the so-called Rabi frequency
$\Omega_{0}=E_{0}\mu/\hbar$~\cite{Scully}. Where, $\hbar$ is the
Planck constant divided by $2\pi$, $\hat{\sigma}_x=|e\rangle\langle
g|+|g\rangle\langle e|$ is the transition operator of the two-level
atom with the ground state $|g\rangle$ and the exciting state
$|e\rangle$, and $\mu$ is the transition matrix element of the
two-level atom.

We consider $k\Delta\ll1$ and $0\ll kx_0\ll\pi/2$, the Hamiltonian
(4) can be approximately written as
\begin{equation}
\hat{H}_{\rm
int}=\hbar\Omega\left(x+x_c\right)(\hat{a}^\dagger+\hat{a})\hat{\sigma}_x
\end{equation}
with the constants $\Omega=k\cos(kx_0)\Omega_{0}$ and
$x_c=\tan(kx_0)/k$. Here, we have used the well-known trigonometric
function $\sin(kx+kx_0)=\cos(kx_0)\sin(kx)+\cos(kx)\sin(kx_0)$ and
neglected the high order of $kx$. Note that, $k\Delta\ll 1$ means
that the range of atomic motion in $x$ direction is much smaller
than the wave length of the cavity mode. The range of $x$ depends on
the initial uncertainty $\Delta$, and the wave packet spread (i.e.,
the diffraction). As mentioned earlier, the diffraction of the atom
is negligible as the duration of the cavity-atom interaction is very
short, i.e., $t\ll m\Delta^2/\hbar$ ($m$ is the mass of atom). Thus,
the value of $x$ is on the order of its initial uncertainty $\Delta$
(e.g., $10~\mu$m), which can be much smaller than the wave length of
cavity mode (about $1$~cm~\cite{HarocheRMP}).

With the interaction (5), the total Hamiltonian of the system can be
written as
\begin{equation}
\hat{H}_p=\frac{p^2}{2m}+\frac{\hbar\omega_{\rm
a}}{2}\hat{\sigma}_z+\hbar\omega_c(\hat{a}^\dagger\hat{a}+\frac{1}{2})+\hbar
\Omega\left(\hat{x}+x_c\right)(\hat{a}^\dagger+\hat{a})\hat{\sigma}_x
\end{equation}
in the Hilbert space of momentum eigenstates. In this space, the
position operator is given by $\hat{x}=i\hbar\partial/\partial p$.
Physically, the first term in the right hand of Eq.~(6) describes
the CM motion of the free atom. The second term describes the atomic
two internal levels (by the Pauli operator
$\hat{\sigma}_z=|e\rangle\langle e|-|g\rangle\langle g|$ and the
transition frequency $\omega_a$). The third term is the free
Hamiltonian of the cavity ground mode. The last term describes the
coupling between the considered three degrees of freedom, i.e., a
position-dependent Jaynes-Cummings interaction. In the rotating
frame defined by $\hat{U}_1=\exp[-ip^2t/(2m\hbar )]$, the
Hamiltonian (6) can be written as
\begin{equation}
\hat{H}_{p}=\frac{\hbar\omega_{\rm
a}}{2}\hat{\sigma}_z+\hbar\omega_c(\hat{a}^\dagger\hat{a}+\frac{1}{2})+\hbar
\Omega(\hat{x}+x_c+\frac{pt}{m})(\hat{a}^\dagger+\hat{a})\hat{\sigma}_x\,
\end{equation}
With such a transform, the free term $p^2/(2m)$ is eliminated.
Considering the atom rapidly crosses the cavity (i.e., the effective
interaction duration $t$ is sufficiently short), there is an impulse
atom-cavity interaction corresponding to the von Neumann
measurement~\cite{AAV,Jozsa}. Thus, $pt/m\rightarrow0$, and the
Hamiltonian (7) reduces to
\begin{equation}
\hat{H}_{p}=\frac{\hbar\omega_{\rm
a}}{2}\hat{\sigma}_z+\hbar\omega_c(\hat{a}^\dagger\hat{a}+\frac{1}{2})+\hbar
\Omega\left(\hat{x}+x_c\right)(\hat{a}^\dagger+\hat{a})\hat{\sigma}_x\,.
\end{equation}
Performing an unitary transformation of $\hat{U}_2=\exp[-i\omega_c
t(\hat{a}^\dagger\hat{a}+1/2)-it\omega_a\hat{\sigma}_z/2]$, the
Hamiltonian (8) further reduces to
\begin{equation}
\hat{H}_{p}=\hbar
\Omega(\hat{x}+x_c)\left(\hat{a}^\dagger\hat{\sigma}_-e^{-i\delta
t}+\hat{a}\hat{\sigma}_+e^{i\delta t}\right)\,
\end{equation}
with the detuning $\delta=\omega_{\rm a}-\omega_c\,$ and the
operators $\hat{\sigma}_-=|g\rangle\langle e|$ and
$\hat{\sigma}_+=|e\rangle\langle g|$. Here, the usual rotating wave
approximation is performed, i.e., the terms relating to the
sum-frequency $\omega_{\rm a}+\omega_c$ have been neglected.

The time-evolution operator for the Hamiltonian (9) can be given by
the Dyson-series:
\begin{equation}
\begin{array}{l}
\hat{U}_{\rm evol}=1+\left(\frac{-i}{\hbar}\right)\int_0^t\hat{H}_p(t_1)dt_1\\
\\
\,\,\,\,\,\,\,\,\,\,\,\,\,\,\,\,\,\,
+\left(\frac{-i}{\hbar}\right)^2\int_0^t\hat{H}_p(t_1)
\int_0^{t_1} \hat{H}_p(t_2)dt_2dt_1\\
\\
\,\,\,\,\,\,\,\,\,\,\,\,\,\,\,\,\,\,+\cdots\,.
\end{array}
\end{equation}
Under the conditions of large detuning: $\Omega\ll\delta$, the above
time-evolution operator can be approximately written as
\begin{equation}
\hat{U}_{\rm evol}\approx e^{-\frac{i}{\hbar}\hat{H}_{\rm eff} t}
\end{equation}
with the effective Hamiltonian $\hat{H}_{\rm
eff}=(\hbar\Omega^2/\delta)(\hat{x}+x_c)^2(\hat{a}^\dagger\hat{a}\hat{\sigma}_z+|e\rangle\langle
e|)$. Considering the cavity is in the vacuum state $|0\rangle$,
i.e., $\hat{a}^\dagger\hat{a}|0\rangle=0$, the effective Hamiltonian
reduces to
\begin{equation}
\hat{H}_{\rm eff}=\hbar
g_0\left(\frac{1}{x_c}\hat{x}+1\right)^2|e\rangle\langle e|\,
\end{equation}
with $g_0=(\Omega x_c)^2/\delta$. This Hamiltonian just describes a
position-dependent vacuum Rabi splitting~\cite{Harochevacuum}, and
the parameter $g_0$ describes the coupling strength between the
internal and external motions of atom. Numerically, considering the
wave length $\lambda=1$~cm of the cavity mode and the Rabi frequency
$\Omega_0/2\pi=10~$KHz~\cite{HarocheRabi}, we have $\Omega
x_c=\Omega_0\sin(k x_0)\approx2\pi\times7$~KHz with $kx_0=\pi/4$,
and consequently $g_0\approx2\pi\times0.7$~KHz with $\Omega
x_c/\delta=0.1$. Of course, as detuning $\delta$ increased, the
coupling strength $g_0$ decreased significantly.

\section{The weak value amplification}
In the following, we will show that the vacuum-induced interaction
(12) can generate a small shift to the initial wave packet
$\phi(p)$, and this displacement can be amplified by using the weak
value technique. In the momentum space, the evolved state of atom
can be written as $|\psi\rangle=\hat{U}_{\rm
evol}|G\rangle|S_i\rangle=\int_{-\infty}^{\infty} dp\psi |p\rangle$,
with $|S_i\rangle$ being the initial state of atomic qubit. We
rewrite the initial Gaussian wave function as
$\phi(p)=\phi(\tilde{p})=(2\pi)^{-1/4}\Delta_p^{-1/2}\exp(-\tilde{p}^2)$
with the dimensionless number $\tilde{p}=p\Delta/\hbar$. Then, we
have
\begin{equation}
\psi=e^{g_c|e\rangle\langle e|\frac{\partial}{\partial
\tilde{p}}}e^{ig_c'|e\rangle\langle e|\frac{\partial^2}{\partial
\tilde{p}^2}}\phi(\tilde{p})|i\rangle\,
\end{equation}
by using the relation $\hat{x}=i\hbar\partial/\partial
p=i\Delta\partial/\partial \tilde{p}$. Here, $g_c=2g_0t(\Delta/x_c)$
and $g_c'=g_0t(\Delta/x_c)^2$ are the dimensionless coupling
parameters, and $|i\rangle=\exp\left(-ig_0t|e\rangle\langle
e|\right)|S_i\rangle$. Considering $\Delta\ll x_c$, i.e., $g_c'\ll
g_c$, the state (13) can be approximately written as
\begin{equation}
\psi=e^{g_c|e\rangle\langle e|\frac{\partial}{\partial
\tilde{p}}}\phi(\tilde{p})|i\rangle\,.
\end{equation}

For simplicity, we re-define
$|i\rangle=\alpha|g\rangle+\beta\exp(i\theta)|e\rangle$ being the
initial internal state of the atoms (which can be prepared by the
well-known single qubit operations). Here, $\theta$ is the phase of
the superposition state, and $\alpha$ and $\beta$ are the
superposition coefficients (real number) satisfying the normalized
condition $\alpha^2+\beta^2=1$. Immediately, we have the state
evolution
$\phi(p)|i\rangle\longrightarrow\alpha\phi(p)|g\rangle+\beta
e^{i\theta}\phi(p+\hbar g_c/\Delta)|e\rangle$, and consequently the
expectation value of atom's momentum reads
\begin{equation}
\langle p\rangle=-\beta^2\frac{\hbar
g_c}{\Delta}=-2\beta^2g_c\Delta_p\,.
\end{equation}
This equation means that, the vacuum in cavity $1$ generates a
transverse shift $\langle p\rangle-0=\langle p\rangle$ to the
average momentum of atoms. Because $\beta^2\leq1$, the shift
$\langle p\rangle\rightarrow0$ for a very weak coupling of
$g_c\rightarrow0$. Furthermore, one can easily calculate the
expectation value $\langle x\rangle=0$ of the atomic position. These
results indicate that the weak coupling $g_c$ can generate
significant changes neither on the observable $\langle p\rangle$ nor
$\langle x\rangle$.

We now use the weak value technique to amplify the shifts $\langle
p\rangle$ and $\langle x\rangle$. First, we preform a single-qubit
operation $\hat{U}=\exp\left(-i\eta\hat{\sigma}_x\right)$ to the
state (14) with the controllable parameter $\eta$. Alternatively,
this single-qubit operation can be realized by the classical
resonant light, as it shown in Fig.~1. Consequently, we have the
final state
\begin{equation}
\begin{array}{l}
\psi'=\hat{U}\psi=\hat{U}e^{g_c|e\rangle\langle
e|\frac{\partial}{\partial
\tilde{p}}}\phi(\tilde{p})|i\rangle\\
\\
\,\,\,\,\,\,\,=\hat{U} \left[1+g_c(|e\rangle\langle
e|)\frac{\partial}{\partial
\tilde{p}}+\frac{g_c^2}{2}(|e\rangle\langle
e|)^2\frac{\partial^2}{\partial
\tilde{p}^2}+\cdots\right]\phi(\tilde{p})|i\rangle\,.
\end{array}
\end{equation}
Second, we post-select an eigenstate of the atomic qubit, e.g.
$|g\rangle$, and immediately the external motion of atoms collapses
on the wave function:
\begin{equation}
\psi'_{w}=\langle g|\psi'\rangle=\langle
g|\hat{U}|i\rangle\left(1+g_cA_w\frac{\partial}{\partial
\tilde{p}}+\frac{g_c^2A_w}{2}\frac{\partial^2}{\partial
\tilde{p}^2}+\cdots\right)\phi(\tilde{p})
\end{equation}
with
\begin{equation}
A_w=\frac{\langle g|(\hat{U}|e\rangle\langle e|)|i\rangle}{\langle
g|\hat{U}|i\rangle}\,.
\end{equation}
Here, we have used the relation $(|e\rangle\langle
e|)^n=|e\rangle\langle e|$ with $n=1,2,3,\cdots$. $A_w$ is our weak
value, although it dos not satisfy the standard definement of
Eq.~(1). This will be explained in the Sec. IV.. Physically, the
post-selection of $|g\rangle$ could be realized by the
field-ionization~\cite{HarocheRabi,HarocheCat,HarocheRMP}. Since
$|e\rangle$ and $|g\rangle$ have the different ionization energies,
the ionization is state selective. Supposing the atoms only in
exciting state $|e\rangle$ are effectively ionized by the applied
moderate electric field, and then the exciting state atoms will be
accelerated in $y$ direction and discarded. However, the ground
state atoms will arrive the plate to be finally detected, as it
shown in Fig.~1.

Considering the weak interaction, i.e., $g_c\ll1$ and
$g_c^2|A_w|\ll1$, the wave function (17) can be approximately
written as
\begin{equation}
\begin{array}{l}
\psi_{w}=\frac{\psi'_{w}}{\langle
g|\hat{U}|i\rangle}=\left(1+g_cA_w\frac{\partial}{\partial
\tilde{p}}\right)\phi(\tilde{p})
\\
\\
\,\,\,\,\,\,\,\,=\phi(p)-
\frac{2g_c\Delta}{\hbar}\text{Re}(A_w)p\phi(p)-i\frac{2g_c\Delta}{\hbar}\text{Im}(A_w)p\phi(p)\,.
\end{array}
\end{equation}
Here, the high orders of $g_c$ have been neglected, and
$\text{Re}(A_w)$ and $\text{Im}(A_w)$ are respectively the real and
imaginary parts of $A_w$. With this approximation, the probability
for successfully post-selecting $|g\rangle$ reads $P\approx|\langle
g|\hat{U}|i\rangle|^2$. According to Eq.~(19), we have the
expectation value of momentum:
\begin{equation}
\langle\hat{p}\rangle_w=\int_{-\infty}^{\infty}\psi^*_{w}p\psi_{w}dp\approx
-\hbar \frac{g_c}{\Delta}
\text{Re}(A_w)=-2g_c\Delta_p\text{Re}(A_w)\,.
\end{equation}
This means that, within the post-selected sub-ensemble the shift of
average momentum $\langle p\rangle_w-0=\langle p\rangle_w$ is
proportional to the real part of the weak value. On the other hand,
in the position presentation, the wave function (19) reads
\begin{equation}
\begin{array}{l}
\phi_w=\int_{-\infty}^{\infty}\psi_{w}\langle x|p\rangle
dp\\
\\
\,\,\,\,\,\,\,\, =\frac{1}{\sqrt{2\pi\hbar}}\int_{-\infty}^{\infty}
\phi(p)e^{ipx/\hbar}dp+\frac{1}{\sqrt{2\pi\hbar}}\frac{\hbar
g_cA_w}{\Delta}\int_{-\infty}^{\infty}
e^{ipx/\hbar}\frac{\partial\phi(p)}{\partial p}dp\\
\\
\,\,\,\,\,\,\,\, =(1-i\frac{g_cA_w}{\Delta}x)\phi(x)
\end{array}
\end{equation}
and consequently the expectation value of positions reads
\begin{equation}
\langle x\rangle_w=\int_{-\infty}^{\infty}\phi_w^*x\phi_wdx\approx
\frac{2g_c}{\Delta}\text{Im}(A_w)\int_{-\infty}^{\infty}\phi(x)x^2\phi(x)dx=2g_c\Delta\text{Im}(A_w)\,.
\end{equation}
This indicates that, within the post-selected sub-ensemble the shift
of average position $\langle x\rangle_w-0=\langle x\rangle_w$ is
proportional to the imaginary-part of the weak value.

Due to the single-qubit operations
$\hat{U}|g\rangle=\cos(\eta)|g\rangle-i\sin(\eta)|e\rangle$ and
$\hat{U}|e\rangle=\cos(\eta)|e\rangle-i\sin(\eta)|g\rangle$, our
weak value reads
\begin{equation}
A_w=\frac{\langle g|(\hat{U}|e\rangle\langle e|)|i\rangle}{\langle
g|\hat{U}|i\rangle}=\frac{1}{Ae^{i\vartheta}+1}
\end{equation}
with $A=\alpha\cos(\eta)/[\beta\sin(\eta)]$ and
$\vartheta=(\pi/2)-\theta$. Consequently, we have
\begin{equation}
\text{Re}(A_w)=\frac{1+A\cos(\vartheta)}{A^2+2A\cos(\vartheta)+1}\,
\end{equation}
\begin{equation}
\,\,\text{Im}(A_w)=\frac{-A\sin(\vartheta)}{A^2+2A\cos(\vartheta)+1}\,.
\end{equation}
These values could be as large as we want by properly adjusting the
parameters $A$ and $\vartheta$. For example, if $\cos(\vartheta)=1$
and $A\rightarrow-1$, then
$\text{Re}(A_w)=1/(1+A)\rightarrow\infty$. If $A=-\cos(\vartheta)$
and $\vartheta\rightarrow0$, then
$\text{Im}(A_w)=\cot(\vartheta)\rightarrow\infty$.
With these enlarged weak values, the weak interaction of $g_c$ could
significantly change the transverse CM motions of atoms via the
basic equations:
\begin{equation}
\,\,\,\,\frac{\langle p\rangle_w}{2\Delta_p}\approx
-g_c\text{Re}(A_w)\,,
\end{equation}
\begin{equation}
\frac{\langle x\rangle_w}{2\Delta}\approx g_c\text{Im}(A_w)\,.
\end{equation}
We would like to emphasize that the shifts $\langle p\rangle_w$ and
$\langle x\rangle_w$ can not be infinitely amplified, as the weak
values were obtained under the weak interaction condition of
$g_c^2|A_w|\ll1$. That is, the amplified displacements of average
position and momentum are limited in the regimes of $g_c\langle
p\rangle_w/(2\Delta_p)\ll1$ and $g_c\langle
x\rangle_w/(2\Delta)\ll1$, respectively. Hence, the present
amplification effects are significant just for the weak interaction
of $g_c\rightarrow0$.

There is a cost of WVA. The probability $P\approx|\langle
g|\hat{U}|i\rangle|^2$ for successfully post-selecting $|g\rangle$
decreases rapidly with the increasing $\text{Re}(A_w)$ or
$\text{Im}(A_w)$, so that the more significant amplification needs
more atoms. In the term of metrology, the WVA may be suboptimal for
parameter estimation since many atoms (information) were
discarded~\cite{SNR,Fisher,Knee2}. However, in the practical
experimental systems the discarded atoms may bring also noises into
the final detection. As it pointed by the previous
refs.~\cite{merit0,merit1,merit2,merit3,SNRZ,JordanX}, the WVA can
offer some certain technical advantages, for example, suppressing
the systematic errors~\cite{SNRZ} or avoiding the detectors
saturation~\cite{JordanX}.

In the present system, it would be very difficult to precisely scan
the position- or momentum-distribution of final atoms. Possibly, one
can place two atoms-detectors (such as the hot-wire
ionizers~\cite{Kapitza-Dirac-RMP}) at the symmetrical positions $x$
and $-x$ to estimate the transverse effects of atoms. In the unit
time, the expected atoms-counting in detectors are given by
$\bar{n}_1=NP\int_{x-l/2}^{x+l/2}|\phi_w(x)|^2dx$ and $\bar{n}_2=
NP\int_{-x-l/2}^{-x+l/2}|\phi_w(x)|^2dx$, respectively. $N$ is the
total number of inputted atoms in the unit time, $l<x$ is the
atoms-collecting region of detectors. According to $\bar{n}_1$ and
$\bar{n}_2$, we have
\begin{equation}
\bar{s}=\frac{\bar{n}_1}{\bar{n}_2}-1=\frac{1+2g_c\text{Im}(A_w)\frac{\bar{x}_l}{\Delta}}{1-2g_c\text{Im}(A_w)\frac{\bar{x}_l}{\Delta}}-1
\approx 4g_c\text{Im}(A_w)\frac{\bar{x}_l}{\Delta}
\end{equation}
with
$\bar{x}_l=\int_{x-l/2}^{x+l/2}x\phi^2(x)dx\big/\int_{x-l/2}^{x+l/2}\phi^2(x)dx$.
Above, the high orders of $g_c$ have been neglected, and $\bar{s}$
can be regarded as the signal of atoms transverse shift. We note
that $\bar{n}_1$, $\bar{n}_2$, and consequently $\bar{s}$ are the
expectation values. In practice, the experimental results may take
$n_{i}=\chi \bar{n}_{i}+\delta_{i}^s+\delta_{i}^r$ (with the index
$i=1,2$) and consequently the equation (28) is replaced by
$s=(n_1/n_2)-1$. $\chi$ is the detection efficiency of the
atoms-detectors. There are two kinds of errors in measurements,
namely systematic error $\delta_{i}^s$ and random error
$\delta_{i}^r$. Certainly, the WVA does not offer advantages for
suppressing the random error since the inputted atoms were reduced
by the post-selection~\cite{SNRZ}. However, it can be found that the
WVA is very useful for suppressing the systematic error which is
proportional to the number of atoms, i.e.,
$\delta_{i}^s=\delta_0\bar{n}_i$ with $\delta_0$ being a small
uncertainty coefficient. This systematic error arises perhaps
because of the unsteady detection efficiency of the atom detector,
the uncertain location of the detector, etc.

\section{Discussion}
Here, we give a brief discussion on the physical meaningful of the
WVA. In the original work of AAV~\cite{AAV}, there are two SG
devices. The first one is used to generate a weak coupling between
the spin and orbit of electron, and the second one is arranged to
preform the post-selection of the electron's spin states. The
present weak measurement processing is similar to that of AAV. The
cavity $1$ plays an atomic SG device to implement the coupling
between the internal qubit and the external CM orbital motion of
atom. The cavity $2$ acts as the second SG device of AAV for
coherently manipulating the atoms. After the cavity $1$ the atom is
in the state (14), which can be written as the standard form of
$\psi\approx\phi(p)|i\rangle-ig_c\hat{A}\hat{P}\phi(p)|i\rangle $,
with $\hat{A}=|e\rangle\langle e|$ and
$\hat{P}=i\hbar\partial/\partial p$. Using the orthonormal
eigenstates $|g\rangle$ and $|e\rangle$ of the two-levels atom,
$\psi$ can be further written as:
\begin{equation}
\psi=(|g\rangle\langle g|+|e\rangle\langle e|)\psi=\langle
g|i\rangle\phi(p,A_g)|g\rangle+\langle
e|i\rangle\phi(p,A_e)|e\rangle\,.
\end{equation}
Here, $A_g=\langle g|\hat{A}|i\rangle/\langle g|i\rangle$,
$A_e=\langle e|\hat{A}|i\rangle/\langle e|i\rangle$,
$\phi(p,A_g)=(1-ig_cA_g\hat{P})\phi(p)$, and
$\phi(p,A_e)=(1-ig_cA_e\hat{P})\phi(p)$.

Obviously, Eq.~(29) represses an entangled state. If the internal
state $|g\rangle$ is measured, then the external motion of atom
collapses on the wave function $\phi(p,A_g)$; whereas if the state
$|e\rangle$ is measured, the atom collapses on $\phi(p,A_e)$. These
measurements preformed on the qubit are just the well-known
projective measurements $\hat{P}_g=|g\rangle\langle g|$ and
$\hat{P}_e=|e\rangle\langle e|$. And the outcomes of $A_g$ and $A_e$
can be regarded as the weak values since they take the same form of
Eq.~(1). However, it can be found that $A_g=0$ and $A_e=1$ because
$\hat{A}=|e\rangle\langle e|$, so that they can not realize the
desirable amplification functions, whatever the initial state
$|i\rangle$ is. We note that, $A_g$ and $A_e$ are both real. Hence,
applying directly the projective measurements to the state (29) can
not yield the effect of positional shifts of atoms, as it mentioned
early.

Comparing to the projective measurement, the weak measurement due to
the post-selection $\hat{P}_f=|f\rangle\langle f|$ is a more general
conception, because the state $|f\rangle$ is beyond the eigenstates
of the system. How can a coherent superposition of the eigenstates
be realized? In AAV's proposal, the desired post-selection is
implemented by the second SG device. It couples the spin to the
$y$-directional orbital motion of electron (the third degree of
freedom of electron). And consequently one can select the
$y$-directional motions (via the strong measurement) to realize a
post-selection of the superposition state of spin (see, e.g., the
ref.~\cite{PRD} which discussed detailedly the AAV's idea). In the
recent optics experiments~\cite{RMP}, the post-selection is realized
by a polarizer which is oriented at a certain angle and then selects
the desirable superposition state of polarization of light.

Here, the cavity $2$ together with the ionization electrodes just
realized an operation $\hat{P}'_f=|g\rangle\langle
g|\hat{U}=|g\rangle\langle f|$ to the state (29). And the weak value
(18) can be written as the standard form of
\begin{equation}
A_w=\frac{\langle g|\hat{U}\hat{A}|i\rangle}{\langle
g|\hat{U}|i\rangle}=\frac{\langle f|\hat{A}|i\rangle}{\langle
f|i\rangle}
\end{equation}
with $\langle f|=\langle g|\hat{U}$. This weak value can be as large
as we want, such as $\text{Im}(A_w)\neq0$. Physically, the present
weak value can be regarded as an outcome of the coherent operation
$\hat{U}$. It can be found that the standard post-selection also
implies the coherent operations, by writing
$\hat{P}_f=|f\rangle\langle f|=\hat{R}|g\rangle\langle
g|\hat{R}^\dagger$ with the unitary evolution operator $\hat{R}$ and
the eigenstate $|g\rangle$ of any systems.

\section{Conclusion}
In this theoretical work, we have shown that a vacuum microwave
cavity can shift the neutral atoms to move transversely. This
non-classical effect is due to the vacuum-induced coupling between
the internal and external motions of free atoms, i.e., a
position-dependent vacuum Rabi splitting. We further showed that the
present effect could be amplified by the weak value technique. After
the atom-cavity coupling, we preformed a single-qubit rotation on
the atomic internal states and consequently post-selected an
internal eigenstate (strong measurement). Then, we obtained a weak
value which was used to amplify the vacuum-induced shift of the
average position or momentum of atoms. Technically, the present WVA
could offers advantages in the practical experiment systems for
observing the weak transverse effect of atoms, such as suppressing
the systematic error of detectors. Physically, our WVA is a
quantum-mechanical effect due to the necessary single-qubit
operation. Finally, we hope the present studies could encourage the
further studies on the weak measurements and cavity-QED.

{\bf Acknowledgements}: This work was partly supported by the
National Natural Science Foundation of China Grants No. 11204249.
\\


\vspace{-1cm}
\begin{thebibliography}{99}
%
\bibitem{AAV}Y. Aharonov, D. Z. Albert, and L. Vaidman, Phys. Rev. Lett. {\bf 60}, 1351 (1988).
%
\bibitem{Jozsa}R. Jozsa, Phys. Rev. A {\bf 76}, 044103 (2007).
%
\bibitem{Amplification1}P. B. Dixon, D. J. Starling, A. N. Jordan, and J. C. Howel,
Phys. Rev. Lett. {\bf 102}, 173601 (2009).
%
\bibitem{Amplification2}O. S. Maga$\tilde{\text{n}}$a-Loaiza, M. Mirhosseini, B. Rodenburg,
and R. W. Boyd, Phys. Rev. Lett. {\bf 112}, 200401 (2014).
%
%
\bibitem{Amplification3}S. Pang, J. Dressel, and T. A. Brun, Phys. Rev. Lett. {\bf 113}, 030401 (2014).
%
\bibitem{Amplification4}Y. Susa, Y. Shikano, and A. Hosoya, Phys. Rev. A {\bf 85}, 052110 (2012).
%
\bibitem{Amplification5}C. Simon and E. S. Polzik, Phys. Rev. A {\bf 83}, 040101 (2011).
%
\bibitem{Hall Effect}O. Hosten and P. Kwiat, Science {\bf 319},
787 (2008).
%
\bibitem{three-box}K. Resch, J. Lundeen, and A. Steinberg, Phys. Lett. A {\bf 324}, 125 (2004).
%
\bibitem{C-Cat}T. Denkmayr, H. Geppert, S. Sponar, H. Lemmel, A. Matzkin, J. Tollaksen, and Y. Hasegawa,
Nat. Commun. {\bf 5}, 4492 (2014).
%
\bibitem{Nature-wavefuction}J. S. Lundeen, B. Sutherland, A. Patel, C. Stewart, and C. Bamber, Nature {\bf 474}, 188 (2011).
%
\bibitem{Science-trajectory}S. Kocsis, B. Braverman, S. Ravets, M. J. Stevens,
R. P. Mirin, L. K. Shalm, and A. M. Steinberg, Science {\bf 332},
1170 (2011).
%
\bibitem{Superluminal}D. Sokolovski, A. Z. Msezane, and V. R. Shaginyan,
Phys. Rev. A {\bf 71}, 064103 (2005).
%
\bibitem{HD}J. S. Lundeen and A. M. Steinberg, Phys. Rev. Lett. {\bf 102}, 020404 (2009).
%
%
\bibitem{LG}J. Dressel, C. J. Broadbent, J. C. Howell, and A. N. Jordan, Phys. Rev. Lett. {\bf 106}, 040402 (2011).
%
%
\bibitem{Uncertain}L. A. Rozema, A. Darabi, D. H. Mahler, A. Hayat, Y. Soudagar, and A. M. Steinberg,
Phys. Rev. Lett. {\bf 109}, 100404 (2012).
%
\bibitem{correlation}H. M. Wiseman, Phys. Rev. A {\bf 65}, 032111 (2002).
%
\bibitem{RMP}J. Dressel, M. Malik, F. M. Miatto, A. N. Jordan, and R. W. Boyd, Rev. Mod. Phys. {\bf 86}, 307 (2014).
%
\bibitem{Dot}N, S. Williams and A, N. Jordan, Phys. Rev. Lett. {\bf 100}, 026804 (2008).
%
\bibitem{Superconducting}N. Katz, M. Neeley, M. Ansmann, R. C. Bialczak, M. Hofheinz,
E. Lucero, A. O'Connell, H. Wang, A. N. Cleland, J. M. Martinis, and
A. N. Korotkov, Phys. Rev. Lett. {\bf 101}, 200401 (2008).
%
\bibitem{AB}O. Zilberberg, A. Romito, and Y. Gefen, Phys. Rev. Lett. {\bf 106}, 080405 (2011).
%
%
\bibitem{Atom}I. Shomroni, O. Bechler, S. Rosenblum, and B. Dayan, Phys. Rev. Lett. {\bf 111}, 023604 (2013).
%
%
\bibitem{HarocheRabi}M. Brune, F. Schmidt-Kaler, A. Maali, J. Dreyer, E. Hagley, J. M. Raimond,
and S. Haroche, Phys. Rev. Lett. {\bf 76}, 1800 (1996).
%
\bibitem{HarocheCat}E. Hagley, X. Ma$\hat{\text{1}}$tre,
G. Nogues, C. Wunderlich, M. Brune, J. M. Raimond, and S. Haroche,
Phys. Rev. Lett. {\bf 79}, 1 (1997).
%
\bibitem{HarocheRMP}J. M. Raimond, M. Brune, and S. Haroche,
Rev. Mod. Phys. {\bf 73}, 565 (2001).
%
\bibitem{splitting1983}
J. J. Sanchez-Mondragon, N. B. Narozhny, and J. H. Eberly, Phys.
Rev. Lett. {\bf 51}, 550 (1983).
%
\bibitem{splittingkimble}
A. Boca, R. Miller, K. M. Birnbaum, A. D. Boozer, J. McKeever, and
H. J. Kimble, Phys. Rev. Lett. {\bf 93}, 233603 (2004).
%
\bibitem{splittingNature}
T. Yoshie, A. Scherer, and J. Hendrickson, Nature {\bf 432}, 200
(2004).
%
\bibitem{Kapitza-Dirac-vacuum}
K. An, Y. -T. Chough, and S. -H. Youn, Phys. Rev. A {\bf 62}, 023819
(2000).
%
\bibitem{Kapitza-Dirac-0}
P. L. Gould, G. A. Ruff, and D. E. Pritchard, Phys. Rev. Lett. {\bf
56}, 827 (1986).
%
\bibitem{Kapitza-Dirac-1}
P. J. Martin, B. G. Oldaker, A. H. Miklich, and D. E. Pritchard,
Phys. Rev. Lett. {\bf 60}, 515 (1988).
%
\bibitem{Kapitza-Dirac-2}
E. M. Rasel, M. K. Oberthaler, H. Batelaan, J. Schmiedmayer, and A.
Zeilinger, Phys. Rev. Lett. {\bf 75}, 2633 (1995).
%
\bibitem{Kapitza-Dirac-RMP}
A. D. Cronin, J. Schmiedmayer, and D. E. Pritchard, Rev. Mod. Phys.
{\bf 81}, 1051 (2009).
%
\bibitem{Scully}M. O. Scully and M. S. Zubairy, \emph{Quantum Optics}, Cambridge University press, 1997.
%
\bibitem{Harochevacuum}S. Haroche, M. Brune and J. M. Raimond,
Europhys. Lett. {\bf 14}, 19 (1991).
%
\bibitem{SNR}C. Ferrie and J. Combes, Phys. Rev. Lett. {\bf 112}, 040406 (2014).
%
\bibitem{Fisher}G. C. Knee and E. M. Gauger,
Phys. Rev. X {\bf 4}, 011032 (2014).
%
\bibitem{Knee2}G. C. Knee, J. Combes, C. Ferrie, and E. M. Gauger,
arXiv: 1410.6252v1.
%
\bibitem{merit0}A. Feizpour, X. Xing, and A. M. Steinberg, Phys. Rev. Lett. {\bf 107}, 133603 (2011).
%
\bibitem{merit1}D. J. Starling, P. B. Dixon, A. N. Jordan, and J. C. Howell, Phys. Rev. A {\bf 80}, 041803 (2009).
%
\bibitem{merit2}G. I. Viza, J. M. Rinc$\acute{\text{o}}$n, G. B. Alves,
A. N. Jordan, and J. C. Howell, Phys. Rev. A {\bf 92}, 032127
(2015).
%
\bibitem{merit3}J. P. Torres and L. J. Salazar-Serrano, arXiv: 1408.1919v1.
%
\bibitem{SNRZ}X. Zhu, Y. Zhang, S. Pang, C. Qiao, Q. Liu, and S. Wu, Phys. Rev. A {\bf 84}, 052111 (2011).
%
\bibitem{JordanX}A. N. Jordan, J. M. Rinc$\acute{\text{o}}$n,
and J. C. Howell, Phys. Rev. X {\bf 4}, 011031 (2014).
%
\bibitem{PRD}I. M. Duck and P. M. Stevenson, Phys. Rev. D {\bf 40}, 2112 (1989).
%


\end{thebibliography}
\end{document}